\documentclass[doublecol]{epl2}
\usepackage{epsfig}
\usepackage{graphicx}
\usepackage{subfigure}

\title{Re-entrant magnetic field induced charge and spin gaps in the coupled dual-chain quasi-one dimensional organic conductor
Perylene$_2$[Pt(mnt)$_2$]}
\shorttitle{Field induced gaps in organic conductor Perylene$_2$[Pt(mnt)$_2]$ } %Insert here a short version of the title if it exceeds 70 characters

\author{L. E. Winter\inst{1} \and J. S. Brooks\inst{1} \and P. Schlottmann\inst{1}  \and M. Almeida \inst{2}  \and S. Benjamin \inst{1}  \and C. Bourbonnais \inst{3}}
\shortauthor{L. E. Winter \etal}

\institute{
  \inst{1} Department of Physics and National High Magnetic Field Laboratory, Florida State University, Tallahassee, Florida 32301 USA\\
  \inst{2} Instituto Tecnol\'{o}gico e Nuclear, Instituto Superior T\'{e}cnico/CFMCUL, Estrada Nacional n$^\circ$ 10, Sacav\'{e}m P-2686-953, Portugal\\
  \inst{3} Regroupement Qu\'{e}b\'{e}cois sur le Mat\'{e}riaux de Pointe, D\'{e}partement de Physique, Universit\'{e} de Sherbrooke, Sherbrooke, Qu\'{e}bec, Canada,
  J1K-2R1
}
\abstract{An inductive method is used to follow the magnetic field-dependent susceptibility of the coupled charge density wave (CDW) and spin-Peierls (SP) ordered state behavior in the dual chain organic conductor Perylene$_2$[Pt(mnt)$_2$].  In addition to the coexisting SP-CDW state phase below 8 K and 20 T, the measurements show that a second spin-gapped phase appears above 20 T that coincides with a field-induced insulating phase.  The results support a strong coupling of the CDW and SP order parameters even in high magnetic fields, and provide new insight into the nature of the magnetic susceptibility of dual-chain spin and charge systems.}

\begin{document}

\maketitle

\section{Introduction}
The quasi-one dimensional organic conductors Per$_2$[M(mnt)$_2$] (where Per = perylene, mnt = maleontirledithioate, and M = Au, Pt, Pd, Ni, Cu, Co, Fe, etc.) exemplify the physics of interacting quasi-one-dimensional conductivity and magnetism \cite{Almeida}.  The subject of this Letter is the behavior of Per$_2$[Pt(mnt)$_2$] in high magnetic fields. Under ambient conditions, the metallic perylene chains undergo a metal-insulator (MI) transition at $T_c$ $\sim$ 8 K, due to the development of a charge density wave (CDW) with  an expected tetramerization of the perylene stacks.  Similarly, the insulating Pt(mnt)$_2$ chains with localized Pt spins (S = $\frac{1}{2}$) undergo a spin-Peierls (SP) transition with dimerization of the Pt(mnt)$_2$ chains \cite{Gama}.  Interestingly, despite the different  wave vectors  of the lattice distortions associated with these two transitions, the $SP_0$ state appears to occur at the same transition temperature as, or as a result of, the $CDW_0$ state \cite{Gama2} (hereafter $SP_0$ and $CDW_0$ refer to the low temperature, low magnetic field state).  At present there is no theoretical work that treats the behavior of this material in terms of separate, but interacting charge and spin chains, although some Kondo-type models have been considered \cite{Bourb, Xavier}.
	
\begin{figure}
\onefigure[scale=0.23]{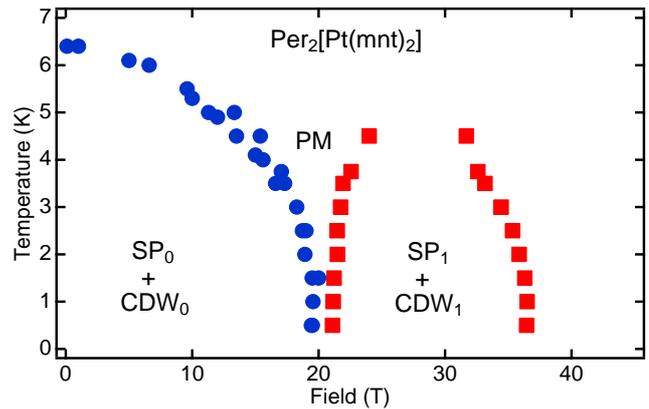}
\caption{Temperature and magnetic field-dependent phase boundaries of Per$_2$[Pt(mnt)$_2$] based on inductive studies (this work). $PM$, $SP_{0}$ + $CDW_{0}$, and $SP_{1}$ + $CDW_{1}$ refer to the paramagnetic metal (normal), low magnetic field, and high magnetic field phases respectively.}
\label{fig.1}
\end{figure}

\begin{figure*}
\begin{center}
\includegraphics[scale=0.23]{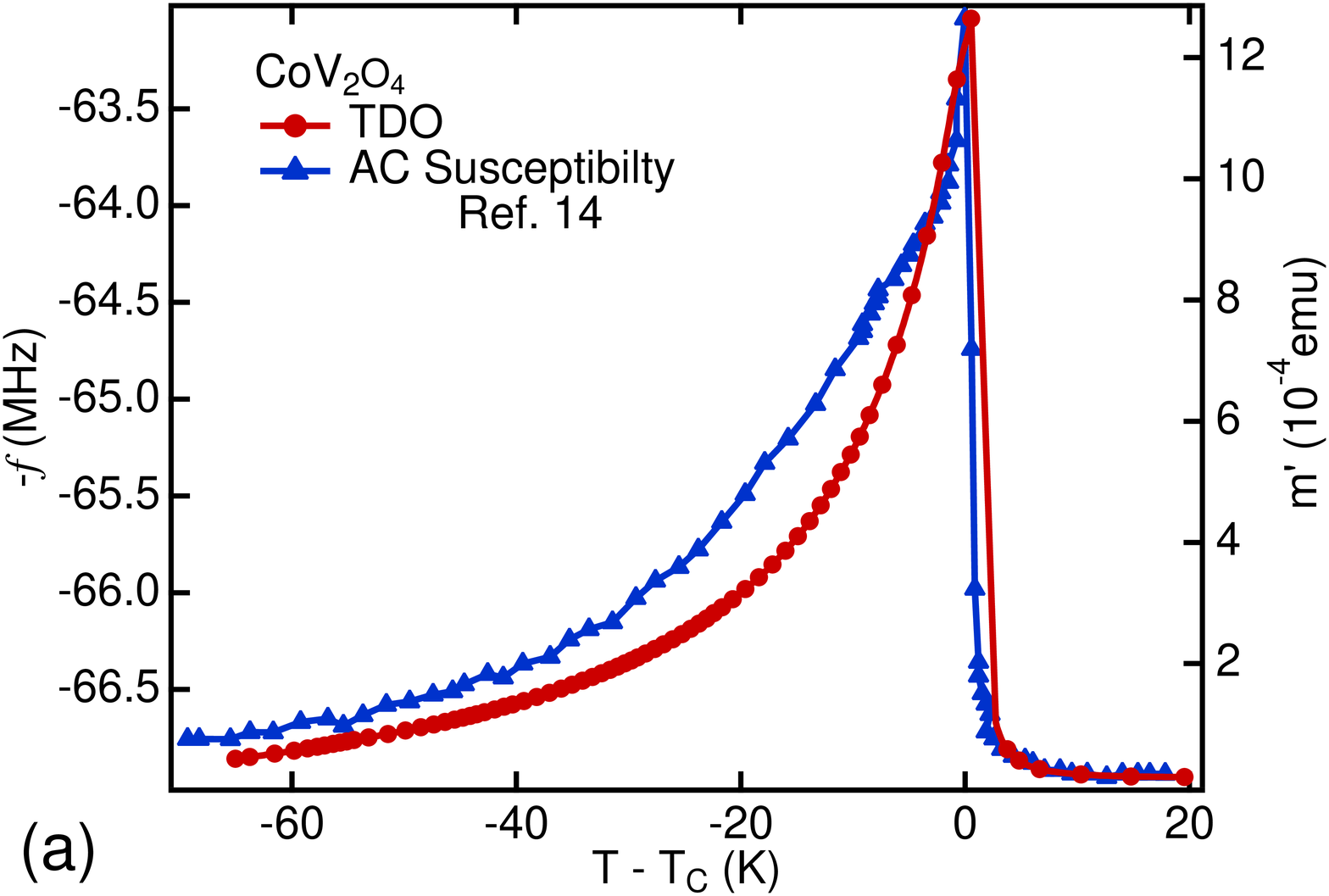}
\includegraphics[scale=0.23]{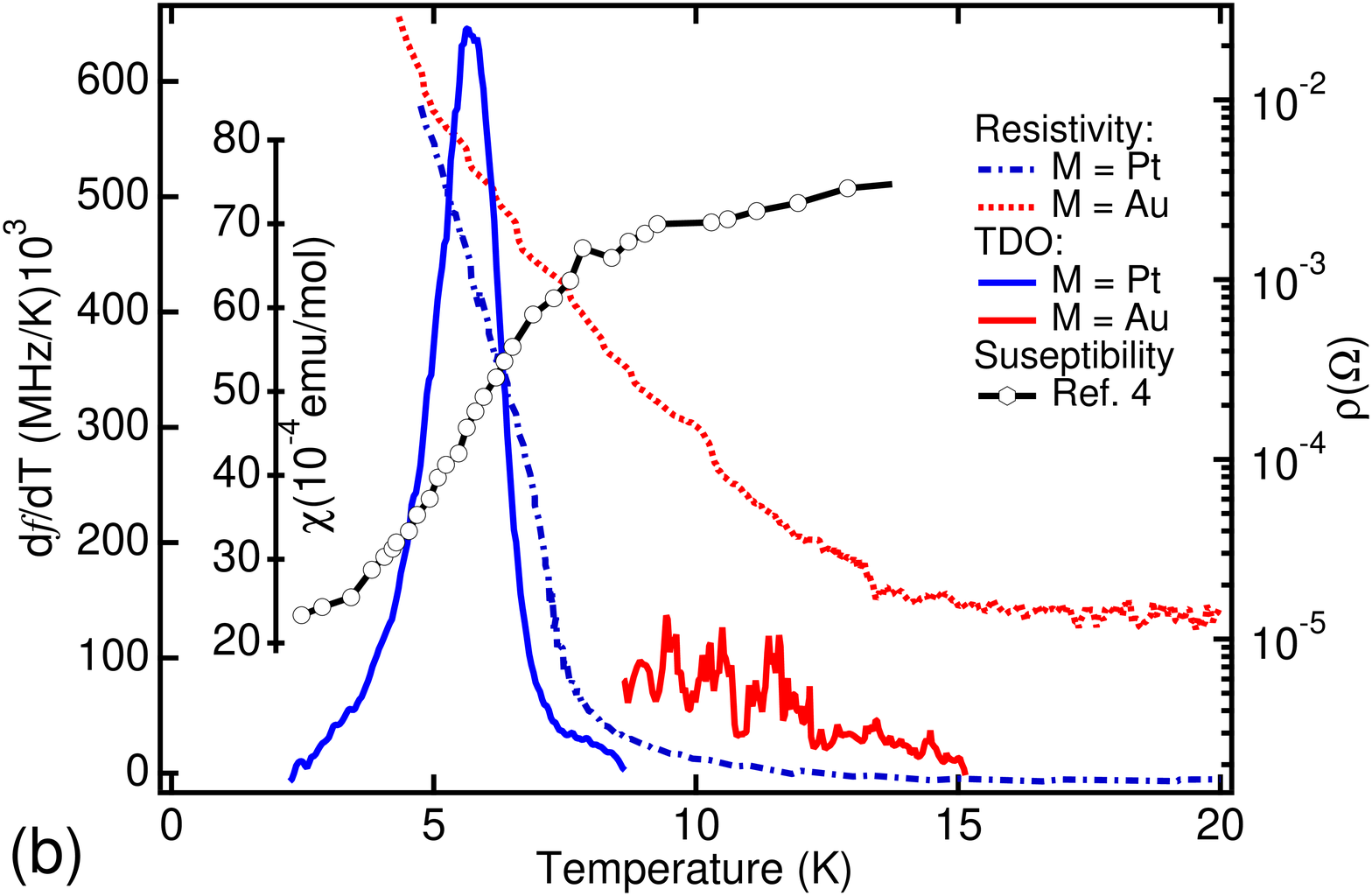}
\caption{a) Comparison of TDO (- $f \sim \chi$) and susceptibility of  CoV$_2$O$_4$ \cite{Huang} in the vicinity of the ferrimagnetic transition at the Curie temperature T$_{C}$ = 152 K.  b) Resistivity, TDO, and susceptibility \cite{Bourb} measurements are compared for Per$_2$[M(mnt)$_2$] where M = Au and Pt.  For M = Pt we can see the temperature derivative of the TDO frequency (note: data taken at 0.1 T), the resistivity $\rho$ , and the susceptibility $\chi$ all confirm the $SP_0$-$CDW_0$ transition below $\sim$ 8K.  For M = Au the resistivity confirms the CDW metal insulator transition at 12 K but no signature is observed in the TDO signal.}
\label{fig.2}
\end{center}
\end{figure*}	
	
Previous electrical transport measurements performed at high magnetic fields on Per$_2$[Pt(mnt)$_2$]  show that in addition to the expected Pauli spin susceptibility-driven suppression of the electrically insulating $CDW_0$ state at high magnetic fields of order 20 T, there is a second, high-magnetic field induced insulating state \cite{Graf} that appears between 20 T and  35 T .  The onset of a magnetic change of phase in Per$_2$[Pt(mnt)$_2$] in this range of fields has also been observed in torque magnetization measurements \cite{Graf, Brooks} and in recent $^1$H NMR experiments \cite{Green, Green2}.  These results support the idea that the order parameters of these two chains remain coupled at least up to 20 T in the main $SP_0$-$CDW_0$ state.  As a consequence, our motivation has been to determine whether this coupling is still present within the field-induced high-field state above 20 T using an inductive ac susceptibility method. In  Fig.~\ref{fig.1} the temperature and magnetic field-dependent phases of Per$_2$[Pt(mnt)$_2$] determined in this work are shown. Here we note that in what follows we describe the regions outside the SP and CDW phase boundaries in terms of a ``normal state" paramagnetic metal.

\section{Experiment}

Measurements were performed using a tunnel diode oscillator (TDO) to obtain results proportional to ac susceptibility \cite{Coffey}.  A TDO is a self-resonating LC circuit (L = inductor, C = capacitor) at frequency $f$, driven by a tunnel diode biased in the negative resistance region.  With a sample in the inductor coil, a change in resonant frequency will occur with a change in magnetic susceptibility $\chi$ and/or skin depth $\delta$ of the material \cite{Coffey, Vannette, Ohmichi}.  Three single crystals with the combined size of roughly 6 x 0.45 x 0.45 mm, obtained as previously described \cite{Afonso}, are contained inside of a small 39 turn coil with an estimated filling factor of 60-75$\%$ and a resonant frequency of $\sim$ 365 MHz.  The long b-axis of the crystals is collinear with the coil and perpendicular to the applied magnetic field (H$\perp$b).  The resonant frequency $f$  is mixed down using a local oscillator at $f_L \sim$ 290 MHz to produce a difference frequency  $\Delta f = f  - f_L \sim$ 75 MHz.  (Since $\Delta f$ only differs from $f$ by a constant, we use $f$ in what follows to simplify the notation.)  To ensure that no signal arose from the superconducting transition of nearby Sn-Pb solder joints in the TDO circuit, all experiments were carried out at fields above 0.1 T. Experiments were first performed up to 16 T in a superconducting magnet, and subsequently in a resistive magnet up to 35 T and to 45 T in the hybrid magnet at the NHMFL in Tallahassee, FL.

Since Per$_2$[Pt(mnt)$_2$] undergoes both a metal insulator transition and the formation of a spin-Peierls ground state, we determined the sensitivity of our method to both $\chi$ and $\delta$ effects (in separate inductive coils similar to the one used in the main experiment). The sensitivity to a purely magnetic transition in an insulator was determined for  CoV$_2$O$_4$ which has a ferrimagnetic transition at the Curie temperature  $T_{C}$ = 152 K, shown in Fig.~\ref{fig.2}a. The TDO clearly follows the magnetic susceptibility obtained by conventional means \cite{Huang}, indicating the change in frequency corresponds to a change in susceptibility. Likewise, the sensitivity to a purely metal-insulator transition (a CDW transition at 12 K) was carried out in non-magnetic Per$_2$[Au(mnt)$_2$], and compared with the signal for Per$_2$[Pt(mnt)$_2$].  The control experiments (both TDO in terms of $df/dT$ and electrical transport in terms of resistivity $\rho$) are shown in Fig.~\ref{fig.2}b. Noting the resistivity in both cases increases exponentially below the respective CDW transition temperatures, only Per$_2$[Pt(mnt)$_2$] shows a significant change in $f$ at a temperature in the vicinity of the resistive transition.  Hence the effects of skin-depth changes due solely to a metal-insulator transition in our measurements are negligible in the TDO response
 \footnote{We have estimated the skin depth of the sample above ($\delta$ = 52 $\mu$m at 8 K) and below ($\delta$ = 200 $\mu$m at 7 K) the metal-insulator transition.  Here $\delta$ =  ($\rho$/$\mu_0 \pi f)^{1/2}$ (where $\rho$ is the resistivity from Gama et al. \cite{Gama2}, $\mu_0$ is the permeability of free space, and $f$ is 365 MHz).   Based on the sample cross-section, by 7 K the sample is fully penetrated by RF and the frequency changes below 7 K must result only from the SP transition.  The change in frequency due to the SP transition can be estimated from skin depth (which allows a determination of the effective sample/coil volume fraction V$_S$/V$_C$) to determine the approximate volume of the sample that contributes to the spin susceptibility around 8 K \cite{Bourb}. Assuming $\chi$ = 0 and $\delta~ >$ sample size below 7 K, $\Delta f \approx -\frac{1}{2}\frac{V_S}{V_C}4\pi \chi_m$  \cite{Vannette} predicts a change of 4.5 MHz.  This is of the same order but higher than that observed in the experiment ($\sim$1.2 MHz).}
compared with changes in the spin-susceptibility. The convention for plotting changes in  -$f$,  shown in Fig.~\ref{fig.2}a and subsequent figures, corresponds to changes in the spin susceptibility $\chi$. (Physically, this means that an increase in $\chi$ corresponds to an increase in $L$, which in turn produces a decrease in $f$. Conversely, as $\chi$ decreases, $f$ will increase.)

\section{Results and analysis}
\begin{figure}
\onefigure[scale=0.23]{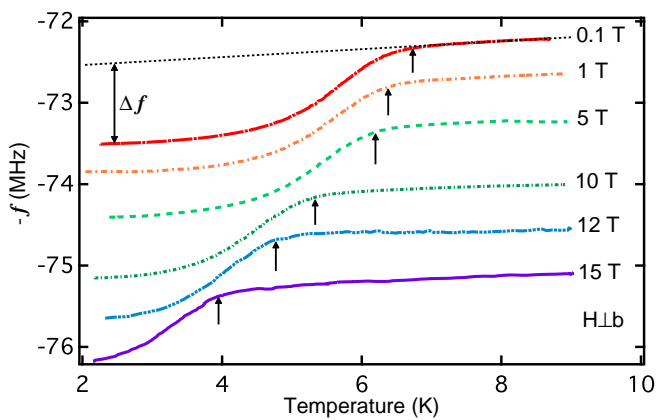}
\caption{The TDO frequency $f$ vs. temperature for fixed magnetic fields (H $\perp$ b) for Per$_2$[Pt(mnt)$_2$].  The dashed line is an extrapolation of the normal (PM) state signal at 0.1 T and $\Delta f$ represents the estimated change in $f$ (i.e. $\Delta f$ = $f_N$ - $f_S$) due solely to the decrease in spin susceptibility as the $SP_0$ state develops. In general, $\Delta f$ will be a function of T and H below the transition. As the field is increased the $SP_0$ transition temperature (arrows) decreases. The data are offset downwards in field to show the transitions clearly.}
\label{fig.3}
\end{figure}

The change in frequency as a function of temperature for fixed magnetic fields up to 15 T is shown in Fig.~\ref{fig.3} for Per$_2$[Pt(mnt)$_2$]. At the lowest field (0.1 T), as the temperature decreases through the $SP_0$-$CDW_0$ transition region, the sample goes from the paramagnetic to spin-singlet state, and the change in frequency reflects the corresponding decrease in susceptibility. With increasing field, the transition temperature decreases. By extrapolating the normal (PM) state signal to low temperatures at each magnetic field we may estimate the change in frequency $\Delta f$ vs. temperature solely due to the decrease in spin susceptibility.

\begin{figure}
\onefigure[scale=0.18]{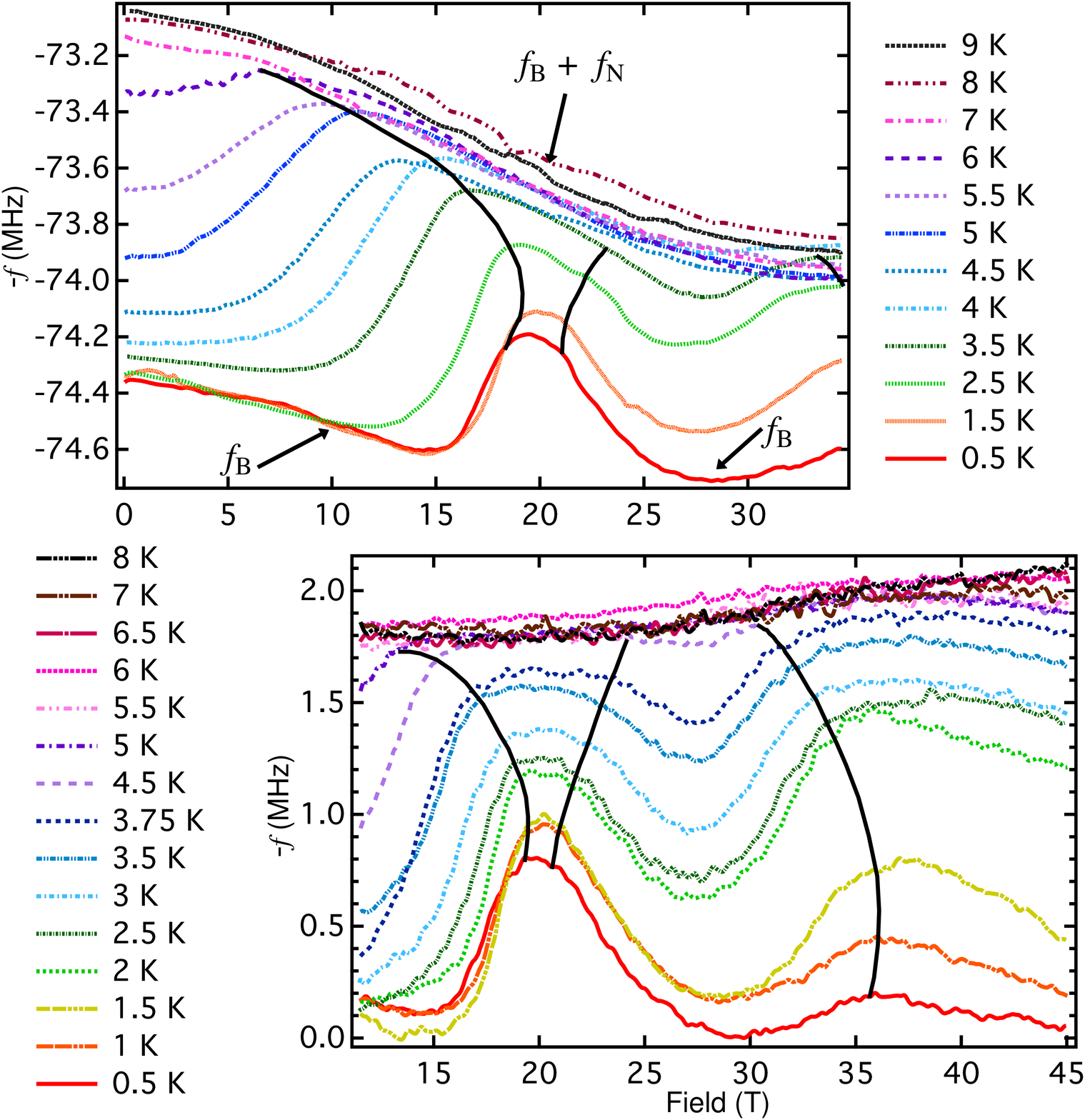}
\caption{Magnetic field dependence (H$\perp$b) of  $f$  for fixed temperatures. $f_N$+$f_B$ refers to regions where the spin gap $\Delta_{1,2}$ is zero (normal state), and $f_B$ refers to regions where the gap is fully developed and only the background signal is present.  Black lines are a guide to the eye to indicate the onsets of the spin gap regions.  Upper panel: measurements up to 35 T. Lower panel: measurements made between 11.5 and 45 T, showing the full extent of the high field phase behavior.  Here, the instrumental TDO background, measured separately over the same field range, has been subtracted.}
\label{fig.4}
\end{figure}

The magnetic field dependence of $f$ up to 35 T at fixed temperatures is shown in Fig.~\ref{fig.4} (upper panel).   Above $\sim$ 8 K, the signal decreases monotonically with magnetic field, and is only weakly dependent on temperature. With decreasing temperature, the change in $f$ is coincident with the onset of the $SP_0$ state, and at fields approaching 20 T $f$ rises, following the onset of the metallic state and the behavior of the phase boundary (Fig.~\ref{fig.1}). However, at higher fields and lower temperatures, there is clearly a second region (between 20 and 35 T) where $f$ decreases again.  This region corresponds to the re-entrant high-resistance phase seen in earlier transport studies and previously described as a field induced charge density wave phase (FICDW) \cite{Graf}. We denote the high-field region between 20 and 35 T, where there are second insulating and reduced spin-susceptibility regions\revision, as $SP_1$ and $CDW_1$ respectively.

In the lower panel of Fig.~\ref{fig.4} the high-magnetic-field phases have been explored between 11.5 and 45 T.  Here the background TDO signal, taken over the same range of temperature and field with the sample removed, has been subtracted from the data, and hence -$f \approx \chi$.  The results show more clearly that the second region between 20 and 35 T is re-entrant to a state with a finite spin susceptibility at higher fields. Under the assumption that above 35 T the system re-enters a paramagnetic state, the high-field susceptibility decreases significantly with decreasing temperature.  We will return to this point in the discussion section below.

\begin{figure*}
\begin{center}
\includegraphics[scale=0.51]{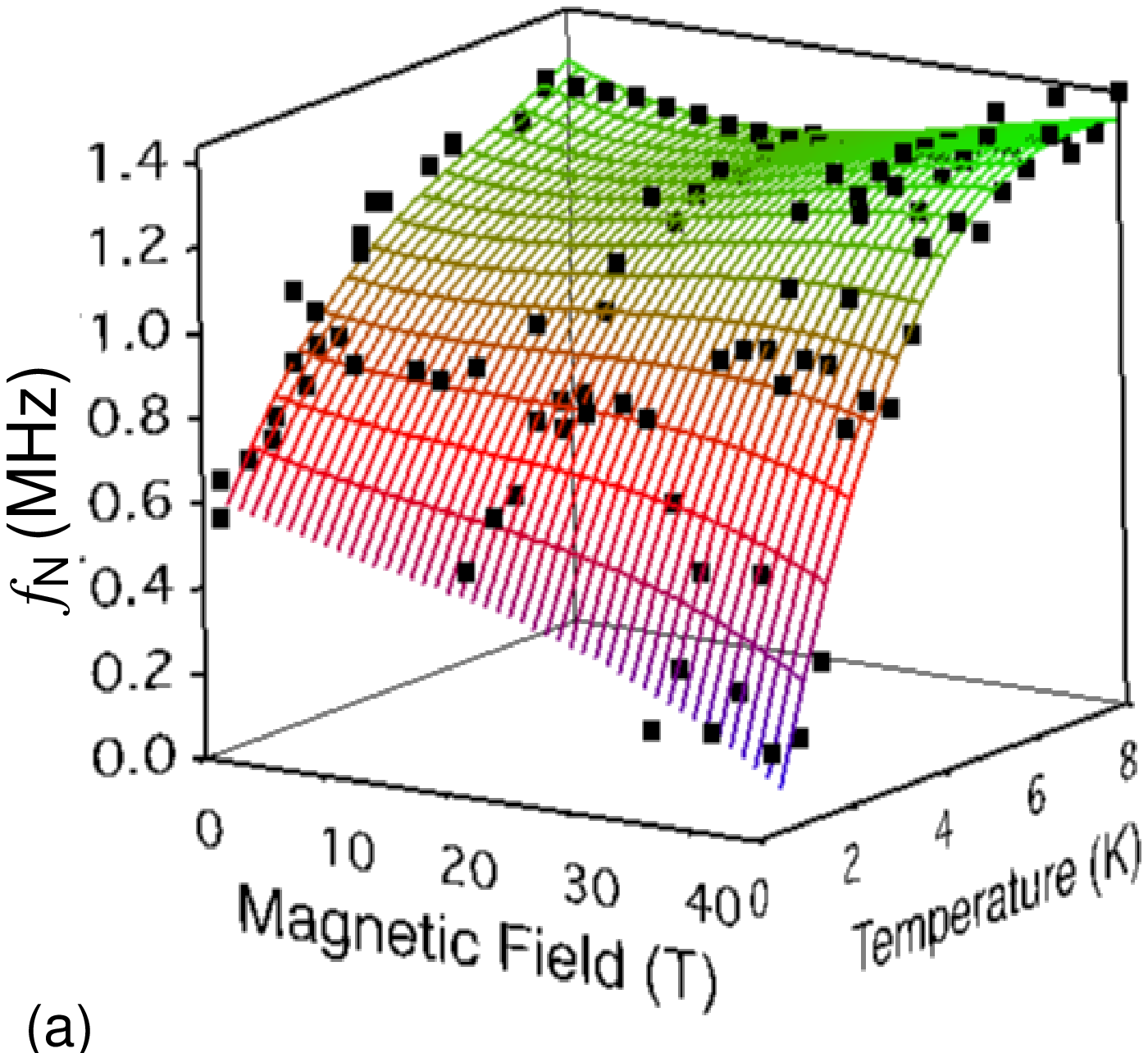}
\includegraphics[scale=0.23]{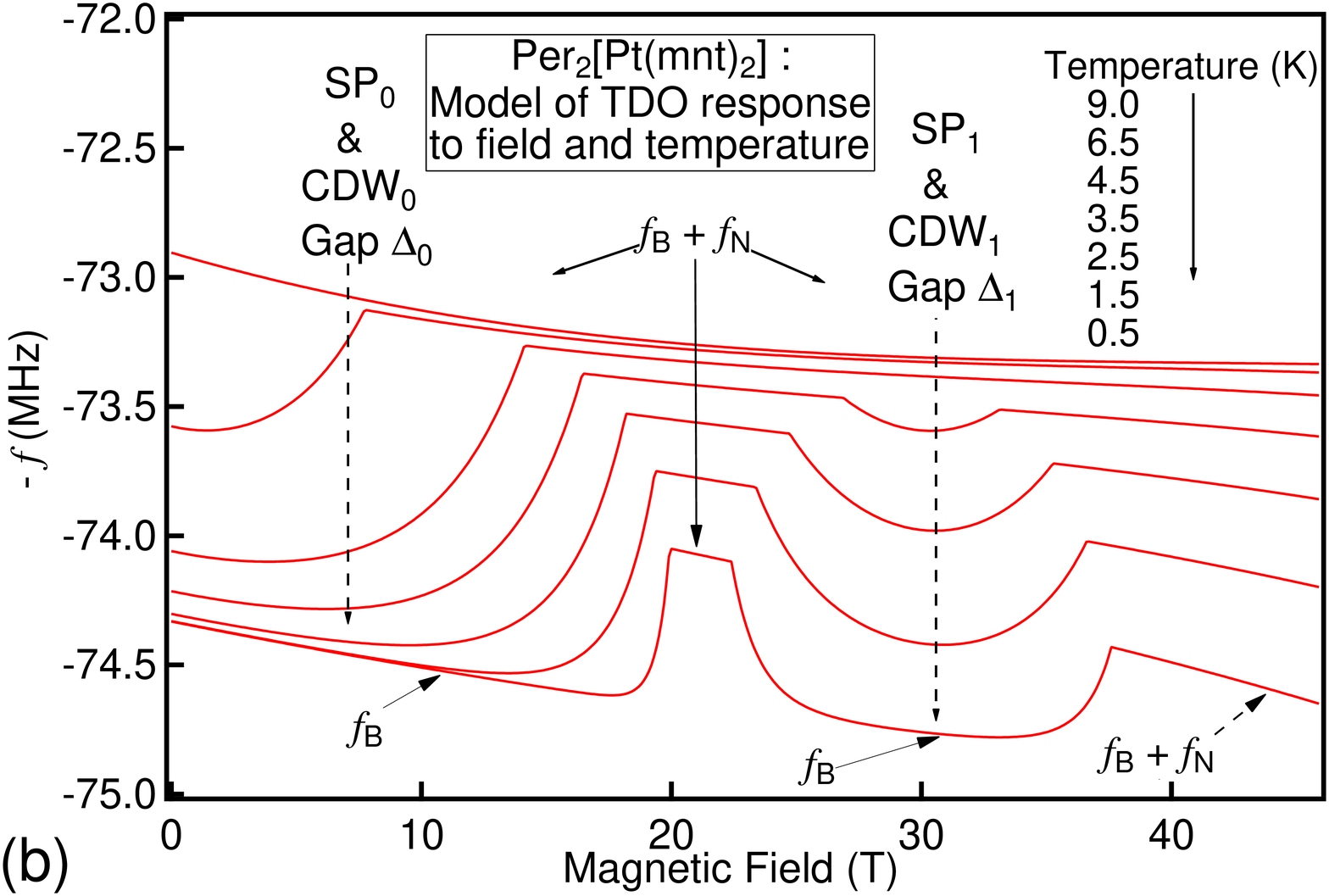}
\caption{a) Estimated field and temperature dependence of the normal state susceptibility $\chi_N$(T,H) (represented as $f_N$).  Solid squares: data from Fig.~\ref{fig.3} derived from $\Delta f$,  and from    Fig.~\ref{fig.4} in regions where $\Delta_{1,0}$ is estimated to be zero. Surface: Two-variable (T and H) polynomial fit to $f_N$. b) Predictions of the simple model for the field and temperature dependent TDO response (compare to Fig.~\ref{fig.4}). Arrows denote the normal state signal $f_N$+$f_B$, the background signal $f_B$, and the regions where the two spin-gaps $\Delta_0$ and $\Delta_1$ develop.}
\label{fig.5}
\end{center}
\end{figure*}

The data for $f$ in Fig.~\ref{fig.4} is the sum of several temperature and magnetic field (T and H) dependent  contributions: i) a background term $f_B$ due to field and temperature dependent spins that remain paramagnetic even in the SP state, plus other instrumental (TDO) background effects; ii) the  contribution from the spin susceptibility $f_S$; and iii) the modification of  $f_S$ when a spin-gap opens. The spin gaps are  T and H dependent, and $\Delta_{0,1}$ represents the spin gap in either the low ($\Delta_0$) or high field ($\Delta_1$) phases.  Assuming a simple exponential model, this corresponds to $f_S  =  f_N$*exp(-$\Delta_{0,1}$ /T) \cite{Green2}, where  $f_N \equiv f_S(\Delta_{0,1} = 0)$ is the signal due to the normal state spin susceptibility in the absence of SP formation. Hence the total signal is  $f = f_S  +  f_B$ where the T and H dependence of all terms is implicit.  The results in Figs.~\ref{fig.3} and~\ref{fig.4} are therefore described as follows. At high temperature above a spin-gapped ground state (where $\Delta_{0,1} = 0$) there is a monotonic dependence of $f$ on T and H.  For lower temperatures, as the SP gap forms,  $f_S$ drops exponentially. When a spin gap is fully formed at low temperatures and low magnetic fields, $f_S$ = 0,   leaving only the background term. From Figs.~\ref{fig.3} and~\ref{fig.4}  we can extract $f_N$  and  $f_B$ by selecting regions of the data that satisfy one of two limiting conditions:  i) when the gap is fully developed,  $f =  f_B$ and the background can be quantified; ii) when the gap is zero,  $f =  f_B + f_N$. Hence as shown in Fig.~\ref{fig.4}, the appearance or disappearance of spin-gapped states will be manifested in changes in the spin susceptibility between these two limits.  This immediately implies that above the main suppression of the $SP_0$-$CDW_0$ ground state with gap $\Delta_0$ at 20 T where the susceptibility again rises towards  the normal state, there is a second, field induced spin-gap $\Delta_1$  between 20 and 35 T where the normal state susceptibility is again driven towards zero, corresponding to the $SP_1$-$CDW_1$ ground state. Above 35 T the normal state susceptibility $f_N$, that has a significant T and H dependence, returns (see following discussion and Fig. 5a).

\section{Discussion}
In light of the above, we have taken selected data from Fig.~\ref{fig.4} where $\Delta_0$ or $\Delta_1$  are  fully developed  to construct an estimate for $f_B$ (taken from data at 0.5 K at low fields and around 35 T), under the simplifying assumption that  $f_B$  is only field dependent. Likewise, we have done the same to find $f_N$ above the SP transition temperature where $\Delta_0$ or $\Delta_1$ = 0 (taken from data above 7 K, from the low temperature changes $\Delta f$ from Fig.~\ref{fig.3}, and in the regions between the spin gaps in the vicinity of 20 T and 45 T where $f$ rises back to a monotonic dependence on H). From these data, we have constructed an extrapolated estimate of the field and temperature dependence of the normal state signal $f_N = f - f_B$ as shown in Fig.~\ref{fig.5}a , which corresponds to the spin susceptibility $\chi (\Delta_{1,2} = 0) = \chi_N$ in the absence of a spin gap. Although $\chi_N$ at low fields bears some resemblance to a Curie-Weiss or Bonner-Fisher type behavior \cite{BF}, neither can describe the functional dependence on T and H: the boundary conditions that our data places require that $\chi_N$ decreases with lower temperatures in high magnetic fields.

Bourbonnais et al. \cite{Bourb2} first noted that the temperature dependent magnetic susceptibility, which shows a maximum above T$_{SP}$, is not driven by quantum spin fluctuations, and is therefore not of a Bonner Fisher type behavior, and that the dimerized SP state is unlikely to form without additional interactions mediated by conduction electrons (RKKY and Kondo-type mechanisms) and the presence of the perylene chains.  For Per$_2$[Pt(mnt)$_2$] the value of the antiferromagnetic coupling constant is estimated to be $J~\sim$ 15 K, and  $T_{SP}^0 ~ = ~25 K $ is the mean field transition temperature at which precursor effects first appear in diffused x-ray diffraction \cite{Henriques}.  In general $J  >> k_BT_{SP}^0$ in a spin-Peierls system, and hence $J$ is too low and this condition is not satisfied\cite{Bourb}, further supporting the necessity of an interaction between the perylene conduction electrons and the Pt chains to stabilize the SP state.

This all implies that a description of the normal state susceptibility and its temperature and magnetic field-dependent behavior will be unusual, and may be coupled to the T and H dependence of the CDW behavior as well. Work based on the specific heat of Per$_2$[Pt(mnt)$_2$]  that considered both a Bonner-Fisher and an Ising description \cite{Bonfait} indicated an Ising description was more likely. The starting point of determining the normal state susceptibility of a spin-Peierls chain is generally that of an Ising or a Heisenberg Hamiltonian with the addition of a lattice energy term, and also a Zeeman energy term in the case of the magnetic field dependence.  There has been theoretical work in this area, for instance by Kl\H{u}mper \cite{Klumper} for a Heisenberg chain, Penson et al.\cite{Penson} for a mixture of Ising and Heisenberg models, and by one of us \cite{Schlottmann} for the Ising chain with and without a spin-Peierls distortion.

In all cases the models described above lead to a reduction of $\chi$ at low temperatures and high magnetic fields in qualitative agreement with Fig.~\ref{fig.5}a, but in the models this is only true above a threshold field, and the models predict the temperature at the maximum in $\chi$ is field dependent. These last two aspects do not correspond to our findings, perhaps because the key factor in any further attempts to model the normal state $\chi$ will involve the inclusion of the interactions with the conducting perylene chain.  However, to our knowledge, there has been no theoretical work that treats the problem of the interaction of two segregated linear chains with different order parameters.

In the present absence of a clear theoretical description of $\chi_N$, we have used the representation  in Fig.~\ref{fig.5}a  to  model  the T and H dependence of $f$  by parameterizing the SP effective spin-gap in the form $\Delta$(H,T) = $\Delta_0$ [1-(H/H$_c$(T))$^2$][1-(T/T$_c$(H))$^2$]  (where $\Delta_0$ = 10 K, H$_c$ = 20 T, and T$_c$ = 8 K). In addition, for the re-entrant spin-gap at high fields we have used a simple semi-circular gap region to approximate a smaller gap opening in the region between 20 and 35 T below 4 K (where $\Delta_1$ = 3 K  and T$_{c1}$ = 4 K). The results of the simple model  $f =  f_B +  f_N$*exp(-$\Delta_{1,2}$/T)  are shown in Fig.~\ref{fig.5}b for  field sweeps at different temperatures, indicating a qualitative description of the experimental data in Fig.~\ref{fig.4}.  In the event that a robust theoretical model for $\chi_N$ were available, it would be possible to use the expression above to derive more precise parameters for the gaps from the experimental data.

For completeness, we use the experimental susceptibility data obtained from Fig.~\ref{fig.4} (lower panel) to obtain the magnetic field dependence of the magnetization at different temperatures by  removing the background $f_B$  and then integrating $f_S$ with respect to magnetic field. The results show a plateau below 20 T which corresponds to M = 0 as expected for the low field $SP_1$ phase. As the field increases above this phase the singlet state starts to break up and the susceptibility (magnetization) starts to increase.  When the second gap starts opening up at the lowest temperatures a second plateau develops that sharpens as the temperature decreases further, and above 35 T there is a rise in the magnetization when the PM state is re-entered. Similar behavior at 0.5 K has been previously reported in the torque magnetization to 45 T \cite{Brooks}; however, in high field NMR studies\cite{Green2}, the temperature (3.1 K) was too high to discern a clear signature of the high field spin gap.

\begin{figure}
\onefigure[scale=0.25]{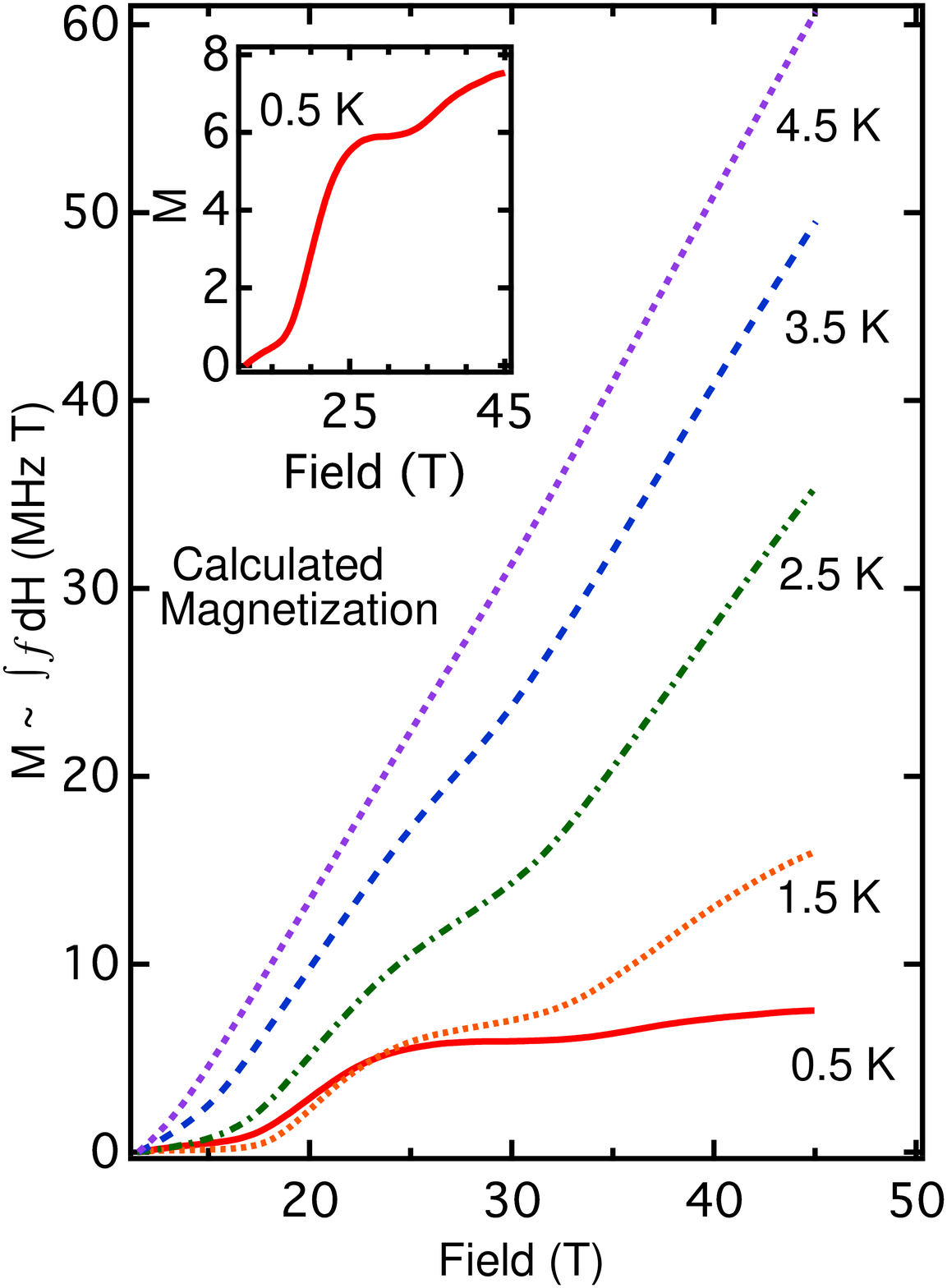}
\caption{Relative magnetization changes vs. magnetic field at different temperatures derived from $f$ in the lower panel of} Fig.~\ref{fig.4}. For the lowest temperatures we observe plateaus that correspond to the two gaps, $\Delta_0$ and $\Delta_1$. The inset shows more clearly the plateau features for 0.5 K. 
\label{fig.6}
\end{figure}

As mentioned previously, the NMR studies on Per$_2$[Pt(mnt)$_2$] suggest the $CDW_0$ state stabilizes the $SP_0$ state at low fields \cite{Green, Green2}. Our data supports this idea since not only does it show similar behavior to that of transport in regards to the low field state, but also in the re-entrant high field state.  If a field induced $CDW_1$ (FICDW) develops, then it makes the possibility of a re-entrant $SP_1$ state as we observe in our data likely, since the re-emergent $CDW_1$ state would re-stabilize the $SP_1$ state.  Our results are also consistent with theoretical work that predicts a FICDW\cite{Lebed, Lebed2} for Per$_2$[Pt(mnt)$_2$]. Another possible explanation given for the observation of the FICDW is due to a CDW-SDW hybridization \cite{Zanchi}. However, with respect to the latter model,  transport studies show that the $CDW_1$ phase is primarily driven by Pauli spin effects since the phase boundaries are not strongly magnetic field direction-dependent, and hence orbital effects may play a lesser role.

The importance of spin and lattice coupling in the magnetic field-dependence of quasi-one-dimensional spin chains is evident in the theoretical work of Vekua \textit{et al.} \cite{Vekua} who predict, for a frustrated spin-1/2 Heisenberg chain with lattice coupling, magnetization plateaus appearing at integer fractions (1/3, 1/2, etc.) of the saturation magnetization.  The results of the magnetization in Fig.~\ref{fig.6} are suggestive in this respect, and it will be important to consider future pulsed magnetic field measurements up to 60 T to explore the possibility of additional plateaus.

\section{Conclusion}		
		The two main results of the present work are as follows.  First, we verify the continued coupling of the CDW and SP order parameters into the high field magnetic-field induced state, where both charge (metal-insulator behavior) and spin (reduced susceptibility) gaps re-emerge. The implication is that the $CDW_0$ is a necessary condition for the formation of the $SP_0$ state in Per$_2$[Pt(mnt)$_2$], and that a FICDW ($CDW_1$) may in turn stabilize the re-entrant spin-gap ($SP_1$) we observe between 20 and 35 T.   Second, we provide a qualitative description of the temperature and magnetic field-dependent normal state magnetic susceptibility of a spin-chain coupled to a conducting chain in the absence of SP formation.  Although theoretical work has treated this problem from the point of view of a single chain with spin and charge degrees of freedom, with some similarities to our results, there is now motivation to extend theoretical models to include the interaction with segregated chains with different order parameters.

\acknowledgments
This work was supported by NSF-DMR 1005293 and the NHMFL is supported by the NSF and the State of Florida. PS is supported by DOE under grant No. DE-FG02-98ER45707.

\end{document}